\documentclass[aip,amssymb,amsmath,reprint]{revtex4-1}
\usepackage{epsfig}
\usepackage[colorlinks,linkcolor=blue,anchorcolor=blue,citecolor=blue,urlcolor=blue,breaklinks=true]{hyperref}
\usepackage{graphics}
\usepackage{color}
\usepackage{graphicx}
\usepackage{dcolumn}
\usepackage{bm}
\usepackage{cases}
\usepackage{appendix}
\usepackage{amsfonts}
\usepackage{epstopdf}
\usepackage{ulem}
\usepackage{subfigure}
\usepackage{bm}
\begin{document}
\address{Department of Physics, Nanjing University, Nanjing 210093, China}
\address{School of Physics and Electronic Engineering, Linyi University, Linyi 276000,China}
\address{Department of Physics, Anhui Normal University, Wuhu 241000, China}
\address{Nanjing Institute of Proton Source Technology, Nanjing 210046, China}
\author{Hao Zhao$^{1, 2}$}
\author{Yong-Long Wang$^{2, 4}$}
 \email{wangyonglong@lyu.edu.cn}
\author{Run Cheng$^{1, 2}$}
\author{Guo-Hua Liang$^{1}$}
\author{Hua Jiang$^{2}$}
\author{Hui Liu$^{1}$}
 \email{liuhui@nju.edu.cn}
\author{Hong-Shi Zong$^{1, 3, 4}$}
\email{zonghs@nju.edu.cn}

\title{Spin polarization of electrons through corrugated surface in magnetic field}
\begin{abstract}
Noninteracting electrons confined to a corrugated surface are investigated in magnetic field, and the associated effective Pauli equation is given analytically by the thin-layer quantization scheme. Interestingly, the Zeeman splitting gaps can be adjusted by curvature, and there is a geometric potential induced by curvature. Further, we discuss the spin-dependent transport properties for confined electrons by numerical calculation. More interestingly, we find that the spin polarization induced by curvature becomes substantial when the incident energy has small value. The results are considerable for a spin transistor with small spin current.
\end{abstract}
\maketitle

As an important link of spintronics, the realization of spin polarization has extensive applications, such as
biology\cite{PNAS.114.2474,PhysRevB.92.115418,PhysRevLett.108.218102,Science.331.894}, materials
science\cite{PhysRevLett.122.127401,JMS.55.250,NanoLett.17.2771,PhysRevLett.120.256601,PhysRevLett.121.120403,PhysRevLett.122.036401}, and optics\cite{PhysRevLett.115.153901}. For these applications, the generation, propagation, and detection of spin polarization in materials are of paramount importance. There are several methods for generating polarized current in various materials. Early experiments demonstrated spin pumping by using optical techniques and nuclear spin system\cite{PhysRevB.15.5780,PhysRevB.13.5484}. More twenty years later, the electrical injection\cite{PhysRevB.65.241306} and strong external bias fields\cite{APL.86.182103} were widely employed to generate spin polarization. Recently, it is found that the strain can be also used to control the spin currents of the junction~\cite{PhysicaB.406.614} and spin splitting in the conduction band~\cite{PhysRevB.94.115131}.\\

\indent  Over the years, the electronic transport in electromagnetic field has been a subject of active studies both theoretically and experimentally\cite{PhysRevB.52.R8646,PhysRevB.72.075312,JAP.88.300,JPSJ.71.543,PhysRevB.47.1466,AIPA.3.072105,EPJAP.68.10201,JPCM.22.253201} owing to application potential. With the advent and development of nanoelectronics, theoretical physicists have tried to study the effects of the geometrical deformation on conductance\cite{PhysRevB.84.195419,IEEETN.6.446,JAP.112.123715}, density of states\cite{PhysRevB.78.115326}, magnetic moment\cite{EPL.111.67004}, magnetoresistance\cite{IJMPB.31.1630016}, resistance\cite{SSC.149.778}, energy level structure\cite{AOP.379.159,PhysicaB.459.88}, persistent currents\cite{PhysRevB.86.035415}, spin current\cite{JPSJ.82.034703} and spin precession\cite{PhysicaE.59.19}. It has been demonstrated that the externally applied electromagnetic field\cite{PhysRevD.100.056014,PhysRevLett.114.076601} and spin-orbit coupling\cite{RevModPhys.76.323,PhysRevB.72.115321,Science.327.1106} are effective methods for spin polarization~\cite{APL.79.3119,APL.78.2184,PhysRevB.66.224412}. To our best knowledge, no significant effect of geometrical deformation on spin polarization has been found~\cite{JPCM.29.135801}. Therefore, in the present letter we will mainly focus on studying how geometric deformation affects spin polarization.\\

\indent In this letter, we will first investigate the effective dynamics for noninteracting electrons confined to a corrugated surface in magnetic field, and then discuss the spin polarization affected by geometrical deformation in a particular case with small transmission probability and small incident energy.

\begin{figure}
  \centering
  \includegraphics[width=0.3\textwidth]{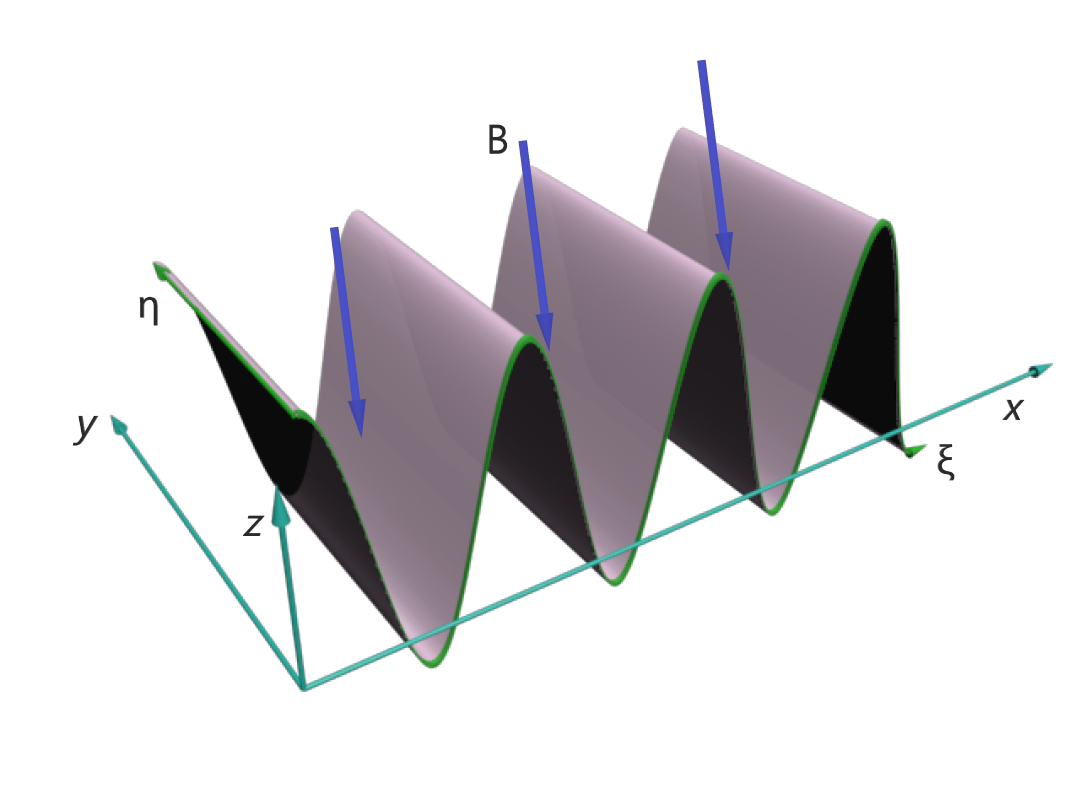}
  \caption{The corrugated surface}\label{fig1}
\end{figure}

\indent Describing an electron with spin in an externally applied magnetic field, the Pauli equation in the usual three-dimensional space is
\begin{equation}\label{equation1}
{\displaystyle \left[{\frac {1}{2m^*}}\left[\left( {\bm{p}} -e {\bm{A}} \right)^{2}-e\hbar { {\bm{\sigma} }}\cdot{\bm{B}} \right]+e\phi
\right]|\psi \rangle
=i\hbar {\frac {\partial }{\partial t}}|\psi \rangle },
\end{equation}
where $\bm{A}$ is a magnetic vector potential, $\phi$ is an electric scalar potential, $m^*$ denotes the effective mass of electron, $e$ is a unit charge, $\hbar$ is a reduced Planck constant, $\bm{\sigma}$ stands for Pauli matrix, $\bm{B}$ describes the magnetic field, and $|\psi\rangle$ is a wave function.
Once an electron is confined to a corrugated surface, which is described by
\begin{eqnarray}
   \bm{r}=\bm{e}_x x+ \bm{e}_y y+\bm{e}_z a cos(\gamma x).\label{eq3}
\end{eqnarray}
In the presence of magnetic field as shown in Fig.\ref{fig1}, the effective Pauli equation~\cite{PhysRevA.90.042117} describing the confined electron can be obtained by using the thin-layer quantization approach~\cite{PhysRevA.23.1982, PhysRevA.25.2893, IEEETED.47.878, PhysRevB.79.235407, PhysicaE.42.1224,PhysRevB.79.235407}, that is
\begin{eqnarray}\
 E \psi =-\frac{1}{2 m^{*}}\{\hbar ^2\partial _{\xi }^2+[\hbar\partial _{\eta }+ ie (B x(\xi)+C)]^2\nonumber\\
  -e\hbar B w\sigma_3\}\psi +V_g\psi, \label{eq12}
\end{eqnarray}
where $\xi$ and $\eta$ stands two tangent vectors of the corrugated surface,
\begin{equation}
  w=\frac{1}{\sqrt{1+a^2 \gamma ^2 sin ^2 (\gamma x)}},
\end{equation}
$\sigma_3$ is a sigma matrix,
\begin{equation}\
  \sigma_3=\left(
    \begin{array}{cc}
      1 &  0 \\
      0 & -1 \\
    \end{array}
  \right),
\end{equation}
and $V_g$ is the well-known geometric potential,
\begin{equation}\label{5}
  V_g=-\frac{\hbar ^2}{8 m^{*}}\frac{[a \gamma ^2 cos(\gamma  x)]^2}{[1+a^2 \gamma ^2 sin^2(\gamma  x)]^3}.
\end{equation}
For simplicity, in the above calculations we have chosen a suitable gauge condition to satisfy the following forms
\begin{equation}\label{6}
A_\xi=0, \\
A_\eta=B[x+q_3 w a \gamma sin(\gamma x)]+ C,\\
A_3=0.
\end{equation}
For the translational invariance of $\eta$, the wave function $\psi$ can be proposed as $\psi =\phi(\xi)  e^{i\eta  k_{\eta }}$, where $k_{\eta}$ is the $\eta$-component of wave vector. For a constant $k_{\eta}$, and by choosing a suitable gauge with $C=-\frac{\hbar
k _{\eta }}{e}$, Eq.\eqref{eq12} can be simplified as
\begin{equation}\label{eq16}
 E \psi =-\frac{1}{2 m^{*}}[\hbar ^2\partial _{\xi }^2-  e^2 B^2 x(\xi)^2
 -e\hbar B w\sigma_3]\psi +V_g\psi.
\end{equation}
It is worthwhile to notice that the Zeeman energy depends on the curvature for $e\hbar B w\sigma_3$ being presented in Eq.~\eqref{eq16}. Specifically, the Zeeman splitting gap will be is considerably affected by the geometric properties of corrugated surface.

\begin{figure}
  \centering
  \includegraphics[width=0.37\textwidth]{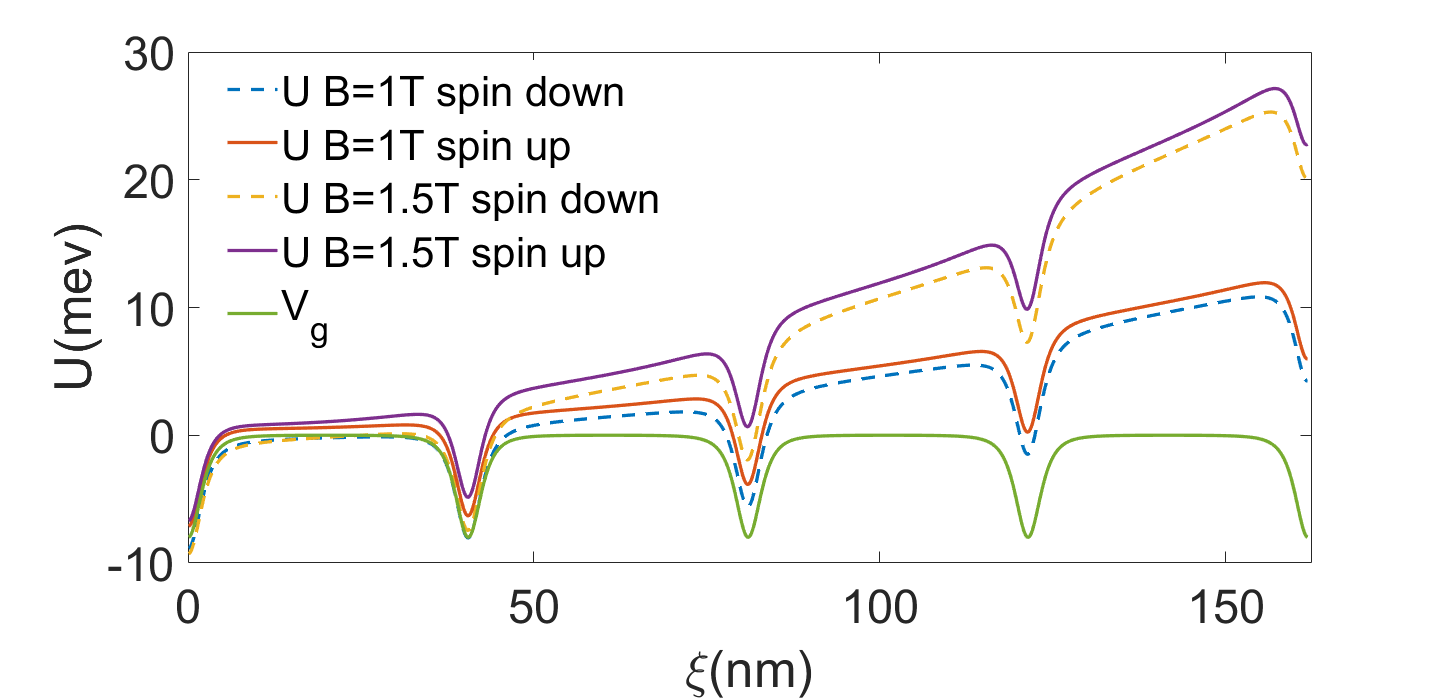}\\
  \caption{The geometrical potential $V_g$ and effective potential $U$ versus $\xi$ when  $a=15nm,\gamma L=4\pi$, where $L$ is the length of the corrugated surface.}\label{fig2}
\end{figure}

In Eq.~\eqref{eq16}, the presence of $e^2 B^2 x(\xi)^2$ shows that the magnetic field acts on the energy spectrum, the presence of $e\hbar B w\sigma_3$ means that the curvature of corrugated surface plays a role of coupling coefficient in the coupling of magnetic field and the normal spin of electron. In what follows, we will focus on the investigation of the curvature acting on the transmission probability and spin polarization. For convenience, the corrugated region is chosen as $0\leq x(\xi)\leq 100nm$, the material is taken as GaAs with $m^*=0.067m_0$ and the leads at both sides have $m^*=m_0$, $m_0$ is the mass of electron. And the terms except the kinetic energy in the right hand side of Eq.~\eqref{eq16} play a role of effective potential, that is
\begin{equation}\label{EffPotential}
U=-\frac{1}{2 m^{*}}[-e^2 B^2 x(\xi)^2-e\hbar B w\sigma_3] +V_g.
\end{equation}
The effective potential is sketched in Fig.~\ref{fig2} for spin up and spin down with the magnetic field $B=1T, 1.5T$. In comparison with the geometric potential, the effective potential $U$ is considerably increased by increasing the external magnetic field. Moreover, the Zeeman splitting gaps between spin up and spin down in general slightly increase by increasing the external magnetic field, and the Zeeman splitting gap is large for large curvature. As a consequence, the Zeeman splitting gap can be controlled by designing the geometrical deformation and introducing magnetic field.

For numerical calculations, the corrugation surface is first divided into $N(N>>1)$
segments, each of which has the same width $\delta\xi$. In the $j$th segment, $\xi_j$ and $\xi_{j+1}$ denote the positions of two sides, respectively, and the effective potential $U$ is approximately taken as a constant $U_{\sigma_3}(\frac{\xi_j+\xi_{j+1}}{2})$. Within the segment, the Eq. (\ref{eq16}) becomes
\begin{eqnarray}\label{13}
  E \psi =-\frac{1}{2 m^{*}}\hbar ^2\partial _{\xi }^2\psi+U_{\sigma_3}(\frac{\xi_j+\xi_{j+1}}{2})\psi.
\end{eqnarray}
According to the principles of quantum mechanics, the wave function $\psi_j(\xi)$ can expressed by
\begin{equation}\label{WaveFunction}
\psi_j(\xi)=\left(
\begin{array}{c}
 \psi_{j+}(\xi) \\
 \psi_{j-}(\xi) \\
\end{array}
\right)
=\left(
\begin{array}{c}
 a_j e^{i k^{+}_{j} \xi}+b_j e^{-i k^{+}_{j} \xi} \\
 c_j e^{i k^{-}_{j} \xi}+d_j e^{-i k^{-}_{j} \xi} \\
\end{array}
\right),
\end{equation}
where $\psi_{j+}(\xi)$ and $\psi_{j-}(\xi)$
stand for the wave function of spin up and spin down, respectively, and $k^{+}_{j}=\frac{\sqrt{2m^{*}[E-U_{j+}]}}{\hbar}$ and $k^{-}_{j}=\frac{\sqrt{2m^{*}[E-U_{j-}]}}{\hbar}$ are two corresponding wave vectors. On account of the boundary continuities of $\psi_j(\xi)$ and $\psi^{'}_j(\xi)$, the $a_j$,$b_j$,$c_j$ and $d_j$ coefficients can be determined by
\begin{equation}\label{16}
  \left(
\begin{array}{c}
 a_j \\
 b_j \\
\end{array}
\right)=\prod_{i=0}^{j-1} M_{i+} \left(
\begin{array}{c}
 a_0 \\
 b_0 \\
\end{array}
\right),
  \left(
\begin{array}{c}
 c_j \\
 d_j \\
\end{array}
\right)=\prod_{i=0}^{j-1} M_{i-} \left(
\begin{array}{c}
 c_0 \\
 d_0 \\
\end{array}
\right),
\end{equation}
where

\begin{eqnarray}
   M_{i+}=\frac{1}{2}\left(
            \begin{array}{cc}
   A_+^+ e^{-iS^+_-\xi_{i+1}} & A_-^+ e^{-iS_+^+\xi_{i+1}} \\
   A_-^+ e^{iS_+^+\xi_{i+1}} & A_+^+ e^{iS_-^+\xi_{i+1}} \\
            \end{array}
          \right),\\
  M_{i-}=\frac{1}{2}\left(
            \begin{array}{cc}
  A_+^- e^{-iS^-_-\xi_{i+1}} & A_-^- e^{-iS_+^-\xi_{i+1}} \\
  A_-^- e^{iS_+^-\xi_{i+1}} & A_+^- e^{iS_+^-\xi_{i+1}} \\
            \end{array}
          \right),
\end{eqnarray}
with $A_{\pm}^+=1\pm\frac{k_i^+}{k_{i+1}^+}$, $A_{\pm}^-=1\pm\frac{k_i^-}{k_{i+1}^-}$, $S^+_{\pm}=k_{i+1}^+\pm k_i^+$ and $S^-_{\pm}=k_{i+1}^-\pm k_i^-$. In the case of $a_0=c_0=1$ and $b_N=d_N=0$, the two coefficients $a_N$ and $c_N$ are obtained as $a_N=\frac{a_0}{M^{+}_{22}}$ and $c_N=\frac{c_0}{M^{-}_{22}}$, respectively, with $M^{+}_{22}=\prod_{i=0}^{N} M_{i+}$
and $M^{-}_{22}=\prod_{i=0}^{N} M_{i-}$. Further, the transmission probabilities for spin up and spin down are obtained as
\begin{equation}
  T_+=1/M^{+}_{22},\quad T_-=1/M^{-}_{22},
\end{equation}
respectively.

On the basis of the above discussions, the transmission probability as a function of the incident energy and the amplitude of corrugation can be numerically calculated in the case of vanishing magnetic field, the results agree well with thsoe known\cite{JPDA.49.295103}. The resonant transmission peaks are substantially deformed by the corrugations. Specifically, the transmission peaks tend to close and aggregate, and the transmission gaps are enlarged by increasing the amplitude of corrugation. As a consequence, the geometric potential can cause the energy spectrum to form band structure.\\
\begin{figure}[htbp]
\centering
  \includegraphics[width=0.48\textwidth]{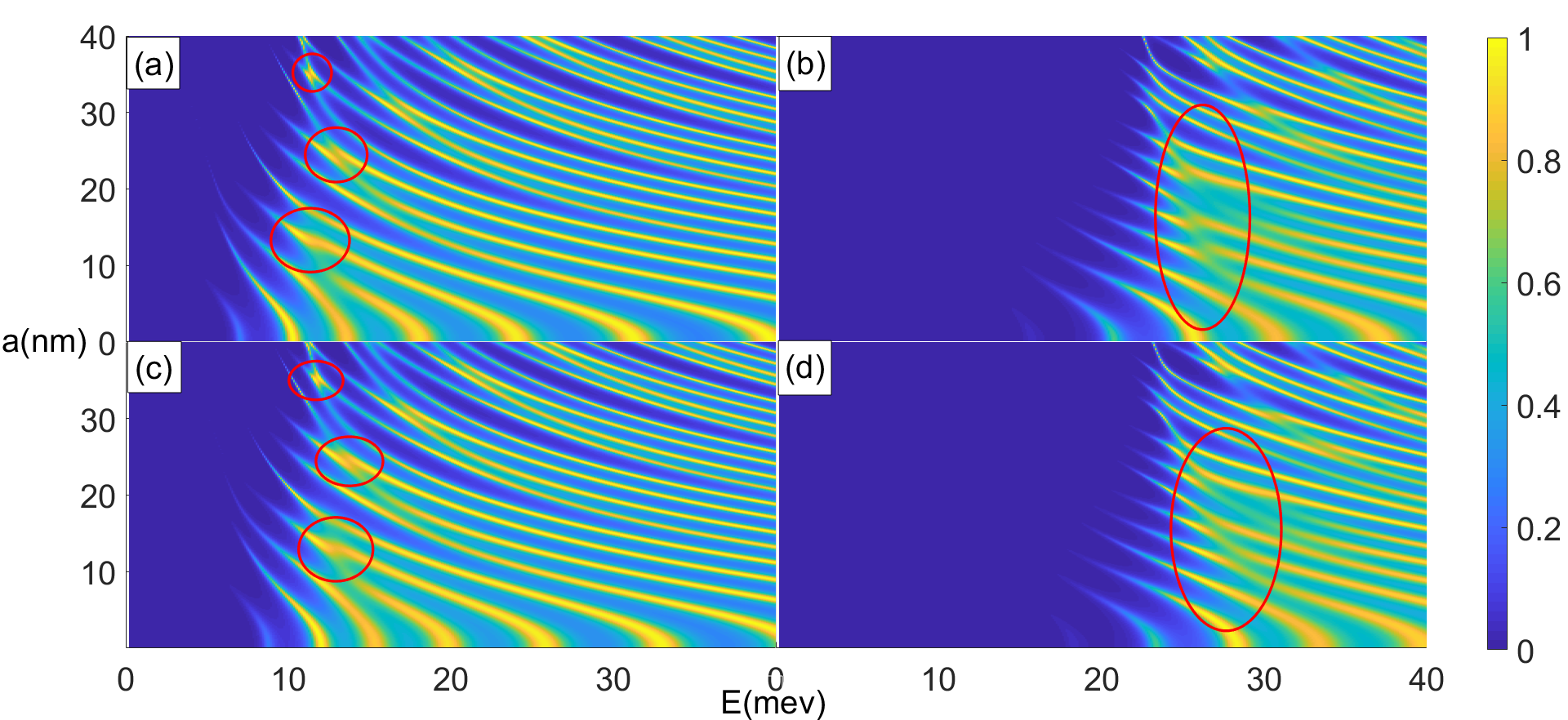}\\
\centering
\caption{The transmission probability versus the incident energy $E$ and the amplitude of corrugation $a$ at $B=1T$ in (a) and (c); $B=1.5T$ in (b) and (d), and $\gamma L= 4\pi$. The top two figures describe  the transmission probability $T_-$ for spin down, the bottom ones describe $T_+$ for spin up.}\label{fig4}
\end{figure}
\begin{figure}[htbp]
  \centering
  \includegraphics[width=0.35\textwidth]{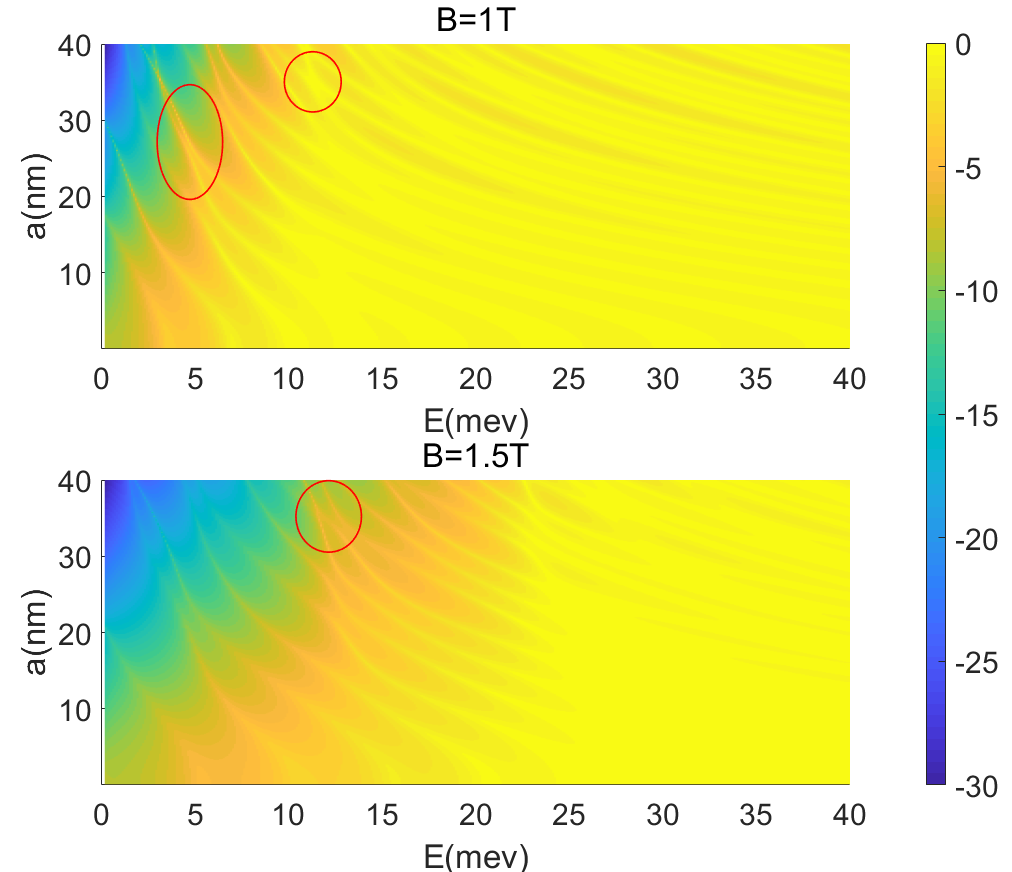}\\
  \caption{The spin down transmission probability $\lg T_-$ versus the incident energy $E$ and the amplitude of corrugation $a$ at $B=1T$ and $B=1.5T$ with $\gamma L= 4\pi$.}\label{fig5}
\end{figure}
\indent As the magnetic field along $z$-axis is externally applied, the transmission probabilities for spin up and spin down are also numerically calculated, the results are sketched in Fig. \ref{fig4}. In comparison with the case of absent magnetic field, these results in the presence of magnetic field are great different. The most significant difference is that the transmission probability has very small value in the area of low incident energy. In other words, the distribution of transmission probabilities in the plane spanned by the incident energy and the amplitude of corrugation, is divided into two domains, a low and a high. With the increase of magnetic field, the low domain expands toward the high domain. The phenomena results from the increase of the effective potential $U$ by increasing magnetic field, shown in Fig.\ref{fig2}. For $U>E$, the order of transmission coefficient can be calculated by $\exp[-\frac{2}{\hbar}\sqrt{2m^*}\int\sqrt{U-E}d\xi]$. The result is described in Fig. \ref{fig5} as the logarithm of spin down transmission coefficient $\lg T_-$ versus $E$ and $a$. It is obvious that the transmission gaps become unconspicuous in the case of strong magnetic field. In addition, the distortion and merging (red circles) of transmission peaks appear in the adjoining area between the low domain and the high domain, implying the competition between the geometric potential and the external magnetic field.\\
\begin{figure}[htbp]
  \centering
  \includegraphics[width=0.35\textwidth]{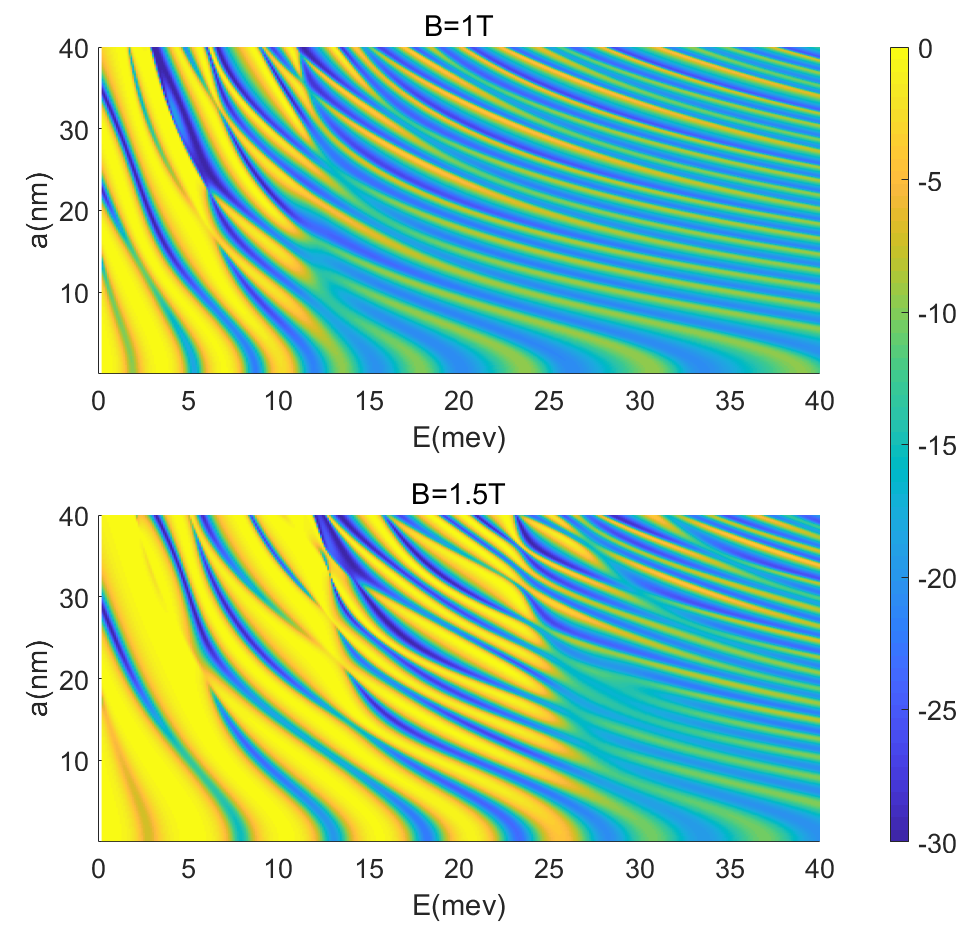}\\

  \caption{The spin polarization $P_T$ versus the incident energy $E$ and the amplitude of corrugation $a$ at $B=1T$ and $B=1.5T$ with $\gamma L= 4\pi$.}\label{fig6}
\end{figure}
\begin{figure}[htbp]
\centering
\includegraphics[width=0.5\textwidth]{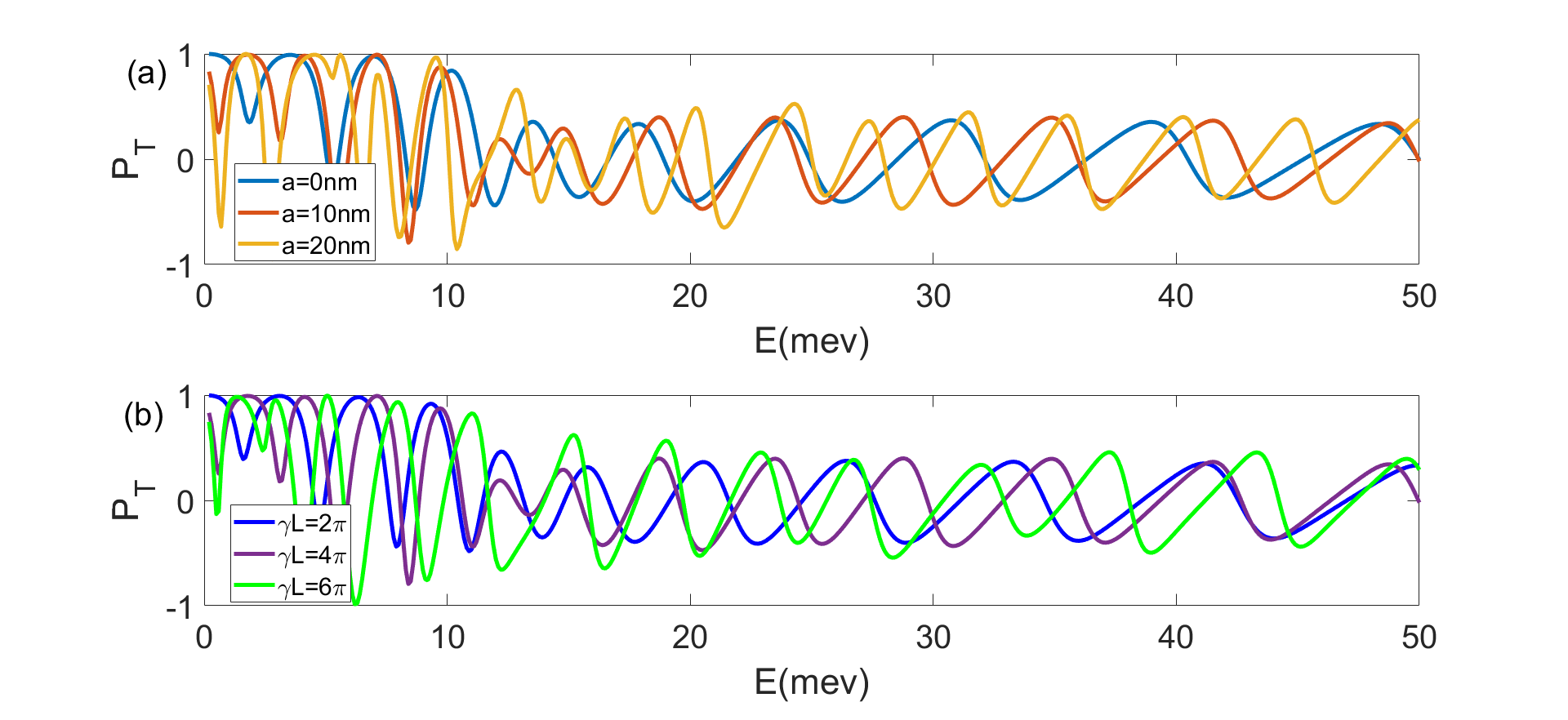}

\centering
\caption{The spin polarization $P_T$ versus the incident energy $E$ with  $B=1T$ for $(a)\gamma L= 4\pi$ and $(b) a=10nm$.}
\hspace{-0.2cm}
\label{fig7}
\end{figure}
\indent In order to investigate the spin polarization affected by the geometrical deformation of curved surface, the spin polarization is defined by
\begin{equation}\label{SpinPolarization}
P_T=\frac{T_+ - T_-}{T_+ + T_-}.
\end{equation}
As described in Fig.\ref{fig6}, $P_T$ has larger values in the area of small $E$ than in that of large $E$, and the domain of $P_T$ with large values will expand toward the area of large incident energy, when the magnetic field becomes large. As shown in Fig.\ref{fig7}, when the amplitude of corrugation increases the transmission probability peaks become sharper and sharper, and the sharpness of the spin polarization becomes better and better. By adding the number of corrugations, the transmission peaks also become sharper, and the associated spin polarization becomes sharper, too. As a conclusion, in the region of small incident energy, the transmission probabilities and the spin polarization are considerably affected by the corrugations. In other words, it is striking that the transmission and spin polarization are substantially influenced by the geometrical properties of corrugated surface in the low incident energy region. These results are differ from the known ones\cite{PhysRevB.66.075331,PhysRevLett.94.246601,JPCM.29.135801}.\\

\indent  In this letter, we have investigated the dynamical properties of electrons confined to a corrugated surface in magnetic field. It is interesting that the Zeeman splitting gap and spin polarization are significantly affected by the corrugations in the area of low incident energy. As an applicable potential, the corrugated surface can be employed to design special spin filter and particular spin transistor for small spin current. In the case of low incident energy, the spin current and spin polarization can be effectively controlled by designing the geometry of curved surface.\\

This work is jointly supported by the National Major state Basic Research and Development of China (Grant No.2016YFE0129300), the National Natural Science Foundation of China (Grants No.11690030,No.11690033, No.11535005, No.61425018). Y.-L. W. was funded by the Natural Science Foundation of Shandong Province of China (Grant No.ZR2017MA010).
\bibliographystyle{apsrev4-1}
\bibliography{reference}
\end{document}